\documentclass[twocolumn,showpacs,preprintnumbers,superscriptaddress,amsmath,amssymb]{revtex4-1}

\usepackage{epsfig}
\usepackage{graphicx}
\usepackage{dcolumn}
\usepackage{bm}
\usepackage{times}
\usepackage{xcolor}

\begin{document}


\title{Suppressing disease spreading by using information diffusion on
multiplex networks}

\author{Wei Wang}
\affiliation{Web Sciences Center, University of Electronic
Science and Technology of China, Chengdu 610054, China}

\affiliation{Big Data Research Center, University of Electronic
Science and Technology of China, Chengdu 610054, China}

\affiliation{Center for Polymer Studies and Department of Physics,
Boston University, Boston, Massachusetts 02215, USA}

\author{Quan-Hui Liu}
\affiliation{Web Sciences Center, University of Electronic
Science and Technology of China, Chengdu 610054, China}
\affiliation{Big Data Research Center, University of Electronic
Science and Technology of China, Chengdu 610054, China}

\author{Shi-Min Cai}
\affiliation{Web Sciences Center, University of Electronic
Science and Technology of China, Chengdu 610054, China}
\affiliation{Big Data Research Center, University of Electronic
Science and Technology of China, Chengdu 610054, China}

\author{Ming Tang}  \email{tangminghan007@gmail.com}
\affiliation{Web Sciences Center, University of Electronic
Science and Technology of China, Chengdu 610054, China}
\affiliation{Big Data Research Center, University of Electronic
Science and Technology of China, Chengdu 610054, China}

\author{Lidia A. Braunstein}
\affiliation{Center for Polymer Studies and Department of Physics,
Boston University, Boston, Massachusetts 02215, USA}
\affiliation{Instituto de Investigaciones F\'{i}sicas de Mar del
Plata (IFIMAR)-Departamento de F\'{i}sica,
Facultad de Ciencias Exactas y Naturales,
Universidad Nacional de Mar del Plata-CONICET,
Funes 3350, (7600) Mar del Plata, Argentina.}

\author{H. Eugene Stanley}
\affiliation{Center for Polymer Studies and Department of Physics,
Boston University, Boston, Massachusetts 02215, USA}

\date{\today}

\begin{abstract}

Although there is always an interplay between the dynamics of
information diffusion and disease spreading, the empirical research on
the systemic coevolution mechanisms connecting these two spreading
dynamics is still lacking. Here we investigate the coevolution
mechanisms and dynamics between information and disease spreading by
utilizing real data and a proposed spreading model on multiplex network.  Our
empirical analysis finds asymmetrical interactions between the
information and disease spreading dynamics. Our results obtained from
both the theoretical framework and extensive stochastic numerical
simulations suggest that an information outbreak can be triggered in a
communication network by its own spreading dynamics or by a disease
outbreak on a contact network, but that the disease threshold is not
affected by information spreading. Our key finding is that there is an
optimal information transmission rate that markedly suppresses the
disease spreading.  We find that the time evolution of the dynamics in
the proposed model qualitatively agrees with the real-world spreading
processes at the optimal information transmission rate.

\end{abstract}

\maketitle

The coevolution dynamics on complex networks has attracted much
attention in recent years, since dynamic processes, ubiquitous in the
real world, are always interacting with each other
\cite{Pastor-Satorras2014,Perc2010}.  In biological spreading dynamics,
two strains of the same disease spread in the same population and
interact through cross immunity \cite{Karrer2010, Sanz2014,Marceau2011}
or mutual reinforcement \cite{Cai2015}.  In social spreading dynamics,
individuals are surrounded by multiple items of information supplied by,
e.g., Facebook, Twitter, and YouTube. These sources of information
compete with each other for the limited attention-span of users, and the
outcome is that only a few items of information survive and become
popular \cite{Gleeson2014,Feng2012}.  Recently scholars have become
aware of the coevolution or interplay between biological and social
spreading dynamics \cite{Funk2010,Funk2010a, Manfredi2013}. When a new
disease enters a population, if individuals who are aware of its
potential spread take preventive measures to protect themselves
\cite{Valdez2012,Alvarez} the disease spreading may be suppressed.  Our
investigation of the intricate interplay between information and disease
spreading is a specific example of disease-behavior systems
\cite{Chris2013}.

Studying the micromechanisms of a disease-behavior system can help us
understand coevolution dynamics and enable us to develop ways of
predicting and controlling the disease spreading \cite{Funk2010}.  In
this effort a number of excellent models \cite{Granell2013,
  Wang2014,Funk2009} have demonstrated the existence of non-trivial
phenomena that differ substantially from those when there is
independent spreading dynamics
\cite{Pastor-Satorras2001,Wang2015,Watts2002,Newman2002,Kuperman2001,
  Castellano2009,Kitsak2010}.  Researchers have demonstrated that the
outbreak of a disease has a metacritical point \cite{Granell2013} that
is associated with information spreading dynamics and multiplex network
topology and that information propagation is promoted by disease
spreading \cite{Wang2014}.  Funk \emph{et al.} found that the disease
threshold is altered once the information and disease evolve
simultaneously \cite{Funk2009}. These models make assumptions about the
coevolution mechanisms of information and disease spreading and do not
demonstrate the interacting mechanisms in real-world systems. Because we
do not understand the microscopic coevolution mechanisms between
information and disease spreading dynamics from real-world
disease-behavior systems, we do not have a systematic understanding of
coevolution dynamics and do not know how to utilize information
diffusion to more effectively suppress the spread of disease.

We present here a systematic investigation of the effects of interacting
mechanisms on the coevolution processes of information and disease
spreading dynamics. We first demonstrate the existence of asymmetrical
interactions between the two dynamics by using real-world data from
information and disease systems to analyze the coevolution. We then
propose an asymmetric spreading dynamic model on multiplex networks to
mimic the coupled spreading dynamics, which will allow us to understand
the coevolution mechanics.  The results, obtained from both the
theoretical analyses and extensive simulations, suggest some interesting
phenomena: the information outbreak can be triggered by its own
spreading dynamics or the disease outbreak, while the disease threshold
is not affected by the information spreading.  Our most important
finding is that there is an optimal information transmission rate at
which the outbreak size of the disease reaches its minimum value, and
the time evolution of the dynamics in the proposed model qualitatively
agrees with the dynamics of real-world spreading.

\section*{Results} \label{sec:real_data}
\textbf{Empirical analysis of real-world coevolution
  data.}
Information about disease can be obtained in many ways, including
face-to-face communication, Facebook, Twitter, and other online
tools. Since the growth of the Internet, search engines have enabled
anyone to obtain instantaneous information about disease.  Patients
seek out and analyze prescriptions using search engines in hopes of
obtaining a means of rapid recovery. Healthy individuals use search
engines to identify protective measures against disease to maintain
their good health.

To examine the coevolution of real-world data about information and
disease, we use weekly synchronously evolving data on information and
disease systems associated with influenza-like illness (ILI) in the US
during an approximate 200-week period from 3 January 2010 to 21
September 2013. The ILI dataset records weekly outpatient visits to
medical facilities, and Google Flu Trends (GFT) dataset keeps track of
week queries in Google search engine about ILI symptoms \cite{http}. The
GFT is used to analyse the occurrence probability of a disease
\cite{Preis2014}.  For simplicity, we assume that the volume of
information about the disease is proportional to the GFT volume because
any individual can use the Google search engine to gain information
about ILI.  For a detailed description of the data see
Ref.~\cite{Preis2014}.

\begin{figure}
\begin{center}
\epsfig{file=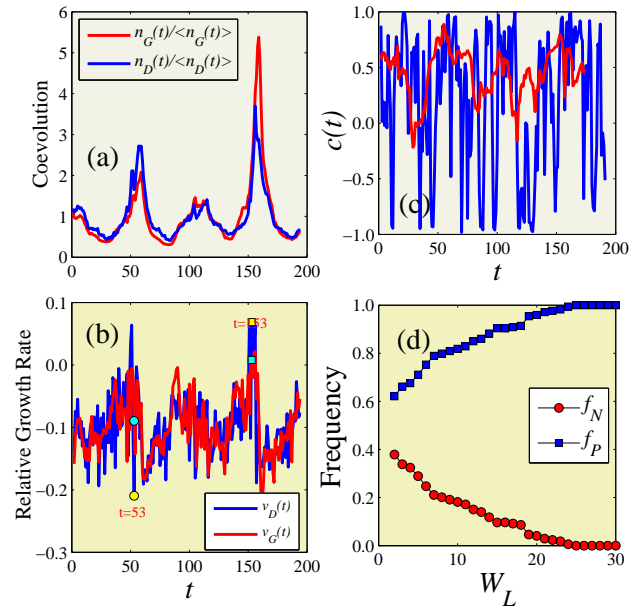,width=1\linewidth}
\caption{(Color online) \textbf{Weekly outpatient visits and Google Flu
    Trends (GFT) of influenza-like illness (ILI) from 3 January 2010 to and 21
      September 2013 in the United States.} (a) The relative number of
      outpatient visits $n_D(t)/\langle n_D(t)\rangle$ (blue dashed
      line) and relative search queries aggregated in GFT
      $n_G(t)/\langle n_G(t)\rangle$ (red solid line) versus $t$, where
      $\langle n_D(t)\rangle=\sum_{t=1}^{t_{\rm max}}n_D(t)/ t_{\rm max}
      $ and $\langle n_G(t)\rangle=\sum_{t=1}^{t_{\rm max}}n_G(t)/
      t_{\rm max} $, and $t_{\rm max}$ is the number of weeks.
      (b) The relative growth rate $v_D(t)$ (blue
  dashed line) and $v_G(t)$ (red solid line) of $n_D(t)$ and $n_G(t)$
  versus $t$, respectively. (c) Cross-correlation $c(t)$ between the two
  time series of $v_G(t)$ and $v_D(t)$ for the given window size $w_l=3$
  (blue dashed line) and $w_l=20$ (red solid line).  (d) The fraction of
  negative correlations $f_P$ (blue squares) and positive correlations
  $f_N$ (red circles) as a function of $w_l$. In (a), $n_G(t)$ and
  $n_D(t)$ are divided their average values respectively. In (b), the
  circles and squares denote the relative growth rate at $t=53$ and
  $153$, respectively.}
\label{fig1}
\end{center}
\end{figure}

Figure~\ref{fig1}(a) shows the real-data time series of
information $n_G(t)$ and disease $n_D(t)$ indicating that
macroscopically the two systems exhibit similar trends and confirming
that the GFT effectively predicts disease spreading
\cite{Preis2014,Ginsberg2009}---although some researchers have expressed
skepticism \cite{Lazer2014}.  To identify the coevolution mechanisms
operating between information and disease spreading, we further
investigate the time series from a microscopic point of
view. Specifically, we study their relative growth rates $v_G(t)$ of
$n_G(t)$ and $v_D(t)$ of $n_D(t)$ (see definitions in Method
Section). Figure~\ref{fig1}(b) shows the evolution of $v_G(t)$ and
$v_D(t)$. Note that the same and opposite growth trends of $v_G(t)$ and
$v_D(t)$ coexist.  For example, at week 53 (week 153), $v_G(53)>0$
[$v_G(153)>0$] and $v_D(53)<0$ [$v_D(153)>0$]. Thus the GFT and ILI show
the opposite (the same) growth trends.

To conceptualize the correlations of the growth trends between the two
dynamics, we analyze the cross-correlations $c(t)$ between the time
series of $v_G(t)$ and $v_D(t)$ for a given window size $w_l$
\cite{Podobnik2008} using the Pearson correlation coefficient $c(t)$
between the two time series $\{v_G(t),v_G(t+1),\cdots,v_G(t+w_l)\}$ and
$\{v_D(t),v_D(t+1),\cdots,v_D(t+w_l)\}$. When $c(t)>0$, the growth rates
of information and disease share the same trend in the time interval
$w_l$. When $c(t)<0$, the information and disease have opposite growth
trends.  Figure~\ref{fig1}(c) shows that the positive and negative
$c(t)$ are uncovered for $w_l=3$ and $w_l=20$, respectively. This may be
because individuals tend to search for disease information when they are
infected or when someone they know is infected, and thus a disease
outbreak promotes the spread of information, i.e., the growth trends of
GFT and ILI will be the same.  When individuals acquire information
about the disease they then take action to protect themselves, and this
causes the growth trends of GFT and ILI to go in opposite directions. We
thus conclude that there are asymmetric interactions between the
dynamics of information and disease spreading, i.e., disease spreading
promotes information spreading, but information spreading suppresses
disease spreading.  Figure~\ref{fig1}(d) plots the fraction of negative
correlations $f_P$ and positive correlations $f_N$ as a function of
$w_l$. The fraction of
positive correlations $f_P$ (negative correlations $f_N$) increases
(decreases) with the $w_l$, since individuals taking measures are
dependent on the timeliness of the information.  Note therefore that
asymmetric interactions can only continue over a short period of time.

\textbf{Coevolution dynamics on multiplex networks.}  We now propose a
novel model based on the coevolution mechanisms in real-world data, i.e., the
asymmetric interactions between information and disease spreading.
Information spreads through communication networks and disease usually
spreads through contact networks. Communication and contact networks
usually have different topologies. To describe the distinct transmission
topologies of the information and disease we use a multiplex network
\cite{Boccalettia2014,Kivela2014,Gao2012,Zhen2015} and construct an
artificial communication-contact coupled network without degree-degree
correlations in intralayers and interlayers.

\begin{figure}
\begin{center}
\epsfig{file=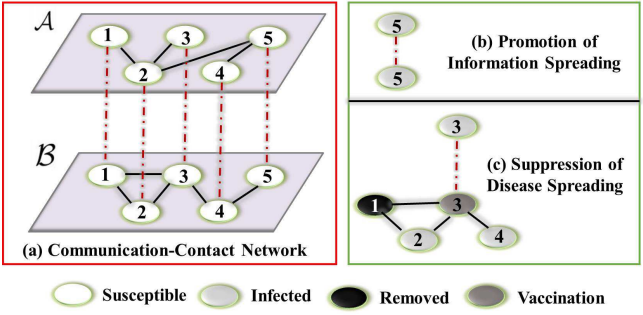,width=1\linewidth}
\caption{(Color online) \textbf{Illustration of asymmetrical mechanisms
    of information and disease on multiplex networks.} (a) A multiplex
  network is used to represent communication and contact networks, which
  are denoted as layer $\mathcal{A}$ and layer $\mathcal{B}$,
  respectively. Each layer has 5 nodes. (b) The promotion of information
  spreading by disease.  If node $5$ on layer $\mathcal{B}$ is infected,
  its counterpart on layer $\mathcal{A}$ becomes informed. (c) The
  suppression of disease spreading by information diffusion. Node $3$ in
  layer $\mathcal{B}$ becomes vaccination only when: (1) its counterpart
  on layer $\mathcal{A}$ is in the informed state and (2) the number of
  its infected neighbors on layer $\mathcal{B}$ is equal to the
  threshold $\phi=2$.}
\label{fig2}
\end{center}
\end{figure}

We generate uncorrelated two-layer networks $\mathcal{A}$ and
$\mathcal{B}$ with degree distributions $P_\mathcal{A}(k_\mathcal{A})$
and $P_\mathcal{B}(k_\mathcal{B})$, where networks $\mathcal{A}$ and
$\mathcal{B}$ represent the communication and contact networks,
respectively. Nodes are individuals and edges are the interactions among
individuals. Each node on layer $\mathcal{A}$ is randomly matched
one-to-one with a node of layer $\mathcal{B}$.  A schematic of the
communication-contact coupled networks is shown in Fig.~\ref{fig2}(a).

Using the analysis results from real-world data, we construct an asymmetric
coevolution information and disease spreading model. In the communication network
(layer $\mathcal{A}$) we use the classic susceptible-infected-recovered
(SIR) epidemiological model \cite{Newman2002,Moreno2002,Serrano2006} to
describe the spreading of information about the disease. Each node can
be in one of three states: susceptible, informed, or recovered. A
susceptible individual has not acquired any information about the
disease, infected (or informed) individuals are aware of the disease and
can transmit their information to their neighbors on the communication
layer, and recovered individuals have the information but do not
transmit it to their neighbors.  At each time step, each informed node
transmits their information to each susceptible neighbor on layer
$\mathcal{A}$ with a probability $\beta_\mathcal{A}$.  The informed node
recovers with a probability $\gamma_\mathcal{A}$. To include the
interacting mechanism between information and disease revealed in the
real-world data analysis, i.e., that disease spreading promotes the
information spreading, we assume that a susceptible node will become
informed when its counterpart in layer $\mathcal{B}$ is infected, as
shown in Fig.~\ref{fig2}(b).

We now introduce a vaccination (V) state into the disease spreading
dynamics on the contact network (layer $\mathcal{B}$) and the model
becomes SIRV \cite{Ruan2012,Buono2015}. The SIR component of the
spreading dynamics is the same as the information spreading on layer
$\mathcal{A}$ and differs only in the infection and recovery rates,
$\beta_\mathcal{B}$ and $\gamma_\mathcal{B}$, respectively. To introduce
the mechanism from our
    real-world data analysis, i.e., that the spread of information
suppresses disease spreading, we assume that an intelligent susceptible
individual on layer $\mathcal{B}$ is vaccinated with probability $p$ (i)
when its counterpart node on layer $\mathcal{A}$ is informed and (ii)
when the number of its neighbors in the infected state is equal to or
greater than a static threshold $\phi$ [see Fig.~\ref{fig2}(c)]. Since
immunization is always expensive, condition (i) means that the
individual must use the communication network to determine the
perniciousness of the disease and condition (ii) means that the
individual will adopt immunization measures only when the probability of
infection is sufficiently high.

We initiate asymmetrical coupled coevolution dynamics by randomly
infecting a tiny fraction of seed nodes on layer $\mathcal{B}$ and
allowing their counterparts on layer $\mathcal{A}$ to become informed.
We set the effective information transmission and disease transmission
rates to be $\lambda_\mathcal{A}=\beta_\mathcal{A}/\gamma_\mathcal{A}$
and $\lambda_\mathcal{B}=\beta_\mathcal{B}/\gamma_\mathcal{B}$,
respectively. Without lack of generality we set
$\gamma_\mathcal{A}=\gamma_\mathcal{B}=1$.  A steady state will be
reached when there are no more nodes in the informed or infected state.

\textbf{Heterogeneous Mean-field theory.}  To quantify the asymmetrical
coevolution dynamics, we develop a heterogeneous mean-field theory. The
outbreak threshold and the fraction of infected or informed nodes in the
final state are the two quantities that control the outcome. For the
information spreading, the densities of susceptible, informed, and
recovered nodes with degree $k_\mathcal{A}$ at time $t$ are denoted by
$s_{k_\mathcal{A}}^\mathcal{A}(t)$,
$\rho_{k_\mathcal{A}}^\mathcal{A}(t)$, and
$r_{k_\mathcal{A}}^\mathcal{A}(t)$, respectively. Analogously, for the
disease spreading, the densities of the susceptible, infected,
recovered, and vaccinated nodes with degree $k_\mathcal{B}$ at time $t$
are denoted by $s_{k_\mathcal{B}}^\mathcal{B}(t)$,
$\rho_{k_\mathcal{B}}^A(t)$, $r_{k_\mathcal{B}}^\mathcal{B}(t)$, and
$v_{k_\mathcal{B}}^B(t)$, respectively.

We first study the time evolution of information spreading on a
communication network, i.e., layer $\mathcal{A}$. The evolution equation
of the susceptible node with degree $k_\mathcal{A}$ on layer
$\mathcal{A}$ can be written
\begin{equation} \label{s_A}
\frac{ds_{k_\mathcal{A}}^\mathcal{A}(t)}{dt}=-s_{k_\mathcal{A}}^\mathcal{A}(t)[\lambda_\mathcal{A}k_\mathcal{A}\Theta_\mathcal{A}(t)+\lambda_\mathcal{B}\langle k_\mathcal{B}\rangle
\Theta_\mathcal{B}(t)],
\end{equation}
where $\langle k_\mathcal{B}\rangle$ is the average degree of layer
$\mathcal{B}$, and $\Theta_\mathcal{A}(t)$ [$\Theta_\mathcal{B}(t)$] is
the probability that a susceptible node connects to an informed neighbor
on uncorrelated layer $\mathcal{A}$ ($\mathcal{B}$) (see details in the
Supporting Information).  The increase in
$\rho_{k_\mathcal{A}}^\mathcal{A}(t)$ is equal to the decrease in
$s_{k_\mathcal{A}}^\mathcal{A}(t)$, and thus the evolution equations for
$\rho_{k_\mathcal{A}}^\mathcal{A}(t)$ and $r_{k_\mathcal{A}}
^\mathcal{A}(t)$ are
\begin{equation} \label{i_A}
\frac{d\rho_{k_\mathcal{A}}^\mathcal{A}(t)}{dt}=s_{k_\mathcal{A}}^\mathcal{A}(t)
[\lambda_\mathcal{A}k_\mathcal{A}\Theta_\mathcal{A}(t)+\lambda_\mathcal{B}\langle
  k_\mathcal{B}\rangle
\Theta_\mathcal{B}(t)]-\rho_{k_\mathcal{A}}^\mathcal{A}(t),
\end{equation}
and
\begin{equation} \label{r_A}
\frac{dr_{k_\mathcal{A}}^\mathcal{A}(t)}{dt}=\rho_{k_\mathcal{A}}^\mathcal{A}(t),
\end{equation}
respectively.

We next investigate the evolution of the disease spreading on layer
$\mathcal{B}$, the contact network. The time evolution equations for the
susceptible, infected, recovered, and vaccinated nodes on layer
$\mathcal{B}$ are
\begin{equation} \label{s_B}
\frac{ds_{k_\mathcal{B}}^\mathcal{B}(t)}{dt}=-\lambda_\mathcal{B}
k_\mathcal{B}
s_{k_\mathcal{B}}^\mathcal{B}(t)\Theta_\mathcal{B}(t)-\Psi(k_\mathcal{B},t),
\end{equation}
\begin{equation} \label{i_B}
\frac{d\rho_{k_\mathcal{B}}^\mathcal{B}(t)}{dt}=\lambda_\mathcal{B}
k_\mathcal{B}
s_{k_\mathcal{B}}^\mathcal{B}(t)\Theta_\mathcal{B}(t)
-\rho_{k_\mathcal{B}}^\mathcal{B}(t),
\end{equation}
\begin{equation} \label{r_B}
\frac{dr_{k_\mathcal{B}}^\mathcal{B}(t)}{dt}=\rho_{k_\mathcal{B}}^\mathcal{B}(t),
\end{equation}
and
\begin{equation} \label{v_B}
\frac{dv_{k_\mathcal{B}}^B(t)}{dt}=\Psi(k_\mathcal{B},t),
\end{equation}
respectively, where $\Psi(k_\mathcal{B},t)$ is the probability that a
susceptible node on layer $\mathcal{B}$ with degree $k_\mathcal{B}$ will
be vaccinated. More details about the Eqs.~(\ref{s_A})--(\ref{v_B}) can
be found in the Supporting Information.

We describe the asymmetrical coevolution dynamics of information and
disease spreading using Eqs.~(\ref{s_A})-(\ref{r_A}) and
(\ref{s_B})-(\ref{v_B}), which allow us to obtain the density of each
distinct state on layer $\mathcal{A}$ and $\mathcal{B}$ at time $t$,
i.e.,
\begin{equation} \label{denstiy_time}
\chi_h(t)=\sum_{k_h}P_h(k_h)\chi_{h_k}^h(t),
\end{equation}
where $h\in\{\mathcal{A},\mathcal{B}\}$ and $\chi\in\{S,I,R,V\}$.  When
$t\rightarrow\infty$, in the steady state, the final sizes of information
and disease systems are $R_\mathcal{A}$ and $R_\mathcal{B}$,
respectively.

Initially only a tiny fraction of nodes on layers $\mathcal{A}$ and
$\mathcal{B}$ are informed or infected, and most are susceptible. Thus
we have $s_{k_\mathcal{A}}^A\approx1$,
$s_{k_\mathcal{B}}^B\approx1$. Linearizing Eqs.~(\ref{i_A}) and
(\ref{i_B}), i.e., neglecting the high order of $\rho_{k_\mathcal{A}}^A$ and
    $\rho_{k_\mathcal{B}}^B$, the critical effective information
transmission probability is
\begin{equation} \label{crit_Inf}
\lambda_c^\mathcal{A}=\frac{1}{\Lambda_C^1},
\end{equation}
where $\Lambda_C^1$ is the maximal eigenvalue of matrix
$$
C=\left(
\begin{array}{ccc}
    C^\mathcal{A} & D^{\mathcal{B}}\\
    0 & C^\mathcal{B}\\
  \end{array}
\right),
$$
$$
C_{k_\mathcal{A},k'_\mathcal{A}}^\mathcal{A}=[\lambda_\mathcal{A}{k_\mathcal{A}}
({k'_\mathcal{A}}-1)P_{\mathcal{A}}(k'_\mathcal{A})]/{\langle
k_{\mathcal{A}}\rangle},
$$
$$
C_{k_\mathcal{B},k'_\mathcal{B}}^\mathcal{B} =
[\lambda_\mathcal{B}{k_\mathcal{B}}({k'_\mathcal{B}}-1)P_{\mathcal{B}}
(k'_\mathcal{B})]/{\langle k_{\mathcal{B}}\rangle},
$$
and
$$
D_{k_\mathcal{B},k'_\mathcal{B}}^\mathcal{B} =
\lambda_\mathcal{B}({k'_\mathcal{B}}-1)P_{\mathcal{B}}(k'_\mathcal{B}),
$$
from which we obtain
\begin{equation} \label{Lambda_c}
\Lambda_C^1=\mathrm{max}\{\Lambda_\mathcal{A}^1,\Lambda_\mathcal{B}^1\},
\end{equation}
where $\Lambda_\mathcal{A}^1$ and $\Lambda_\mathcal{B}^1$ are the
maximal eigenvalues of the adjacent matrix of layers $\mathcal{A}$ and
$\mathcal{B}$, respectively. More details can be found in the Supporting
Information. The critical value $\lambda_c^\mathcal{A}$ separates
information spreading dynamics into local and global information
regions.  When $\lambda_\mathcal{A}\leq\lambda_c^\mathcal{A}$, it is in
the local information region. When
$\lambda_\mathcal{A}>\lambda_c^\mathcal{A}$, it is in the global
information region. In Eq.~(\ref{crit_Inf}) the global information
outbreak condition is correlated only with the topologies of layers
$\mathcal{A}$ and $\mathcal{B}$, i.e., the immunization probability $p$
and threshold $\phi$ do not affect the outbreak of information, but
increasing the degree heterogeneity of layers $\mathcal{A}$ and
$\mathcal{B}$ increases the information outbreak probability.

When $\lambda>\lambda_{c}^\mathcal{A}$, immunization can suppress
disease spreading on subnetwork $\mathcal{B}$, and thus here
immunization process and disease spreading can be treated as competing
processes \cite{Karrer2010}.  Reference~\cite{Karrer2010} demonstrates
that the two competing processes can be treated as one after the other
in the thermodynamic limit.  When the immunization process spreads more
quickly than the disease, it first spreads on layer $\mathcal{B}$ and
then the disease spreads on the residual network (i.e., the network
after immunization). When the disease spreads more quickly than the
immunization, the opposite occurs.  Using
Refs.~\cite{Karrer2010,Wang2014} we find that the immunization
progresses more quickly than the disease, i.e.,
$\lambda_\mathcal{A}\lambda_{\mathcal{B}u}>
\lambda_\mathcal{B}\lambda_{\mathcal{A}u}$, in which
$\lambda_{\mathcal{A}u}=\langle k_\mathcal{A}\rangle/(\langle
k_\mathcal{A}^2\rangle-\langle k_\mathcal{A}\rangle)$ and
$\lambda_{\mathcal{B}u}=\langle k_\mathcal{B}\rangle/(\langle
k_\mathcal{B}^2\rangle-\langle k_\mathcal{B}\rangle)$, which are the
thresholds for the SIR model on a one-layer network \cite{Newman2002},
and $\langle \cdots\rangle$ are the moments of the degree distribution.
Because in many real-world scenarios information spreads more quickly
than disease, we focus on that case. Thus immunization and disease
spreading on layer $\mathcal{B}$ can be treated successively and
separately.  When $\phi=0$, the approximate disease threshold is
\begin{equation} \label{crit_Epi_0}
\lambda_c^\mathcal{B}=
\frac{\langle k_\mathcal{B}\rangle}{(1-V_\mathcal{B})(\langle
k_\mathcal{B}^2\rangle-\langle k_\mathcal{B}\rangle)},
\end{equation}
which is the same as in Ref.~\cite{Wang2014}. In Eq.~(\ref{crit_Epi_0}),
where $V_\mathcal{B}=pQ_\mathcal{A}$, and $Q_\mathcal{A}$ is the final density of the
informed population without disease spreading obtained using link
percolation theory \cite{Newman2002}. From Eq.~(\ref{crit_Epi_0}) we can
see that, as expected, the threshold is bigger than in the SIR model
without vaccination.

When $\phi\geq 1$ we use competing percolation theory to obtain the
approximate disease threshold. The information first spreads on
layer $\mathcal{A}$, and then
    the disease spreads on layer $\mathcal{B}$. Although many nodes on
    layer $\mathcal{A}$ receive the information for large values of
    $\lambda_\mathcal{A}$, the counterparts of those informed nodes
    still cannot be immunized when $\lambda_\mathcal{B}$ is small. This
    is the case because according to the proposed model the susceptible
    nodes that are vaccinated must have authentication from both layers
    $\mathcal{A}$ and $\mathcal{B}$. These informed nodes cannot acquire
    authentication from layer $\mathcal{B}$ when $\lambda_\mathcal{B}$
    is below the disease threshold. Only for large values of
    $\lambda_\mathcal{B}$, these informed nodes can obtain authentication
    simultaneously from layers $\mathcal{A}$ and $\mathcal{B}$. Here the
    immunized nodes are $V_B\approx0$ and thus the approximate disease
threshold is
\begin{equation} \label{crit_Epi_Obove}
\lambda_c^\mathcal{B}=\frac{\langle k_\mathcal{B}\rangle}{\langle
  k_\mathcal{B}^2\rangle-\langle k_\mathcal{B}\rangle},
\end{equation}
which is the same as the outbreak threshold of SIR disease
\cite{Newman2002}, i.e., this kind of information-based immunization
strategy does not affect the disease outbreak threshold, and this
differs from the existing results \cite{Granell2013,Wang2014}.  The
disease threshold is dependent only on the topology of layer
$\mathcal{B}$ and is independent
    of the topology of layer $\mathcal{A}$, the immunization
probability $p$, and the threshold $\phi$.  The asymmetrical coevolution
mechanisms presented in our model may explain why the disease threshold is
not altered in some real-world situations
\cite{Fisman2014,Ali2013,Bermejo2006}.

\textbf{Simulation results.}  We perform extensive stochastic
simulations to study the proposed asymmetrically interacting spreading
dynamics on multiplex networks. In the simulations the network sizes and
average degrees are set at $N_\mathcal{A} =N_\mathcal{B}=10^4$ and
$\langle k_\mathcal{A}\rangle=\langle k_\mathcal{B}\rangle=8$,
respectively. We use the uncorrelated configuration model to generate
layers $\mathcal{A}$ and $\mathcal{B}$ according to the given degree
distributions \cite{Catanzaro2005}. For each multiplex network, we
perform the dynamics $10^4$ times and measure the average final fraction
of information size $R_\mathcal{A}$, disease size $R_\mathcal{B}$, and
immunization size $V_\mathcal{B}$ with five randomly selected seeds in
layer $B$.  We then average these results over 100 network realizations.

To understand the coevolution dynamics of information and disease, we
use Erd\H{o}s-R\'{e}nyi (ER) networks to represent the communication and
contact networks. The degree distributions of layer $\mathcal{A}$ and
layer $\mathcal{B}$ are $P_\mathcal{A}(k_\mathcal{A})=e^{-\langle
  k_\mathcal{A}\rangle} \langle
k_\mathcal{A}\rangle^{k_\mathcal{A}}/k_\mathcal{A}!$ and
$P_\mathcal{B}(k_\mathcal{B})=e^{-\langle k_\mathcal{B}\rangle} \langle
k_\mathcal{B}\rangle^{k_\mathcal{B}}/k_\mathcal{B}! $, respectively.

\begin{figure}
\begin{center}
\epsfig{file=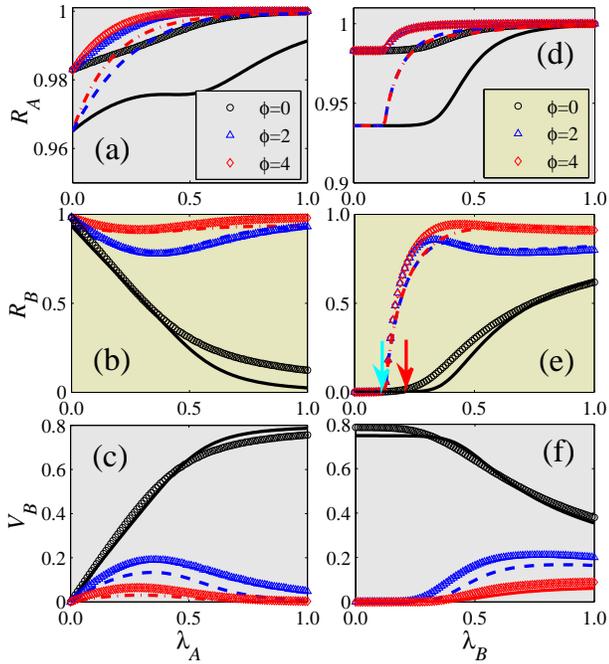,width=1\linewidth}
\caption{(Color online) \textbf{With
      immunization thresholds $\phi$ being the parameter of interest,
      the final sizes of information, disease and vaccination on two
      layer ER-ER multiplex networks.} (a) The final information size
  $R_\mathcal{A}$, (b) the final disease size $R_\mathcal{B}$, and (c)
  the final vaccination size $V_\mathcal{B}$ versus information
  transmission rate $\lambda_\mathcal{A}$ for different values of
  immunization threshold $\phi$ with $\lambda_\mathcal{B}=0.5$.  For
  different values of $\phi$, (d) $R_\mathcal{A}$, (e) $R_\mathcal{B}$
  and (f) $V_\mathcal{B}$ as a function of $\lambda_\mathcal{B}$ at
  $\lambda_\mathcal{A}=0.5$.  The symbols represent the simulation
  results and the lines are the theoretical predictions obtained by
  numerically solving Eqs.~(\ref{s_A})-(\ref{r_A}) and
  (\ref{s_B})-(\ref{v_B}).  In (e), the two arrows respectively indicate
  the numerical disease thresholds for $\phi\geq1$ and $\phi=0$, which
  are obtained by observing $\chi$. Other dynamical parameters are set
  to be $\lambda_\mathcal{B}=0.5$ and $p=0.8$.}
\label{fig3}
\end{center}
\end{figure}

Figure~\ref{fig3} shows how the immunization threshold $\phi$ affects
the final information, disease, and vaccination sizes. For the
information spreading on layer $\mathcal{A}$, we find that
$R_\mathcal{A}$ increases with $\lambda_\mathcal{A}$ and
$\lambda_\mathcal{B}$ [see Figs.~\ref{fig3}(a) and (d)]. In addition,
$R_\mathcal{A}$ increases with $\phi$ because the individuals in layer
$\mathcal{B}$ need a large $\phi$ value to guide their immunization
decisions [see Figs.~\ref{fig3}(c) and (f)], which causes
$R_\mathcal{B}$ to increase with $\phi$ [see Figs.~\ref{fig3}(b) and
  (e)].  As a result, the information spreading increases as disease
spreading increases.

\begin{figure}
\begin{center}
\epsfig{file=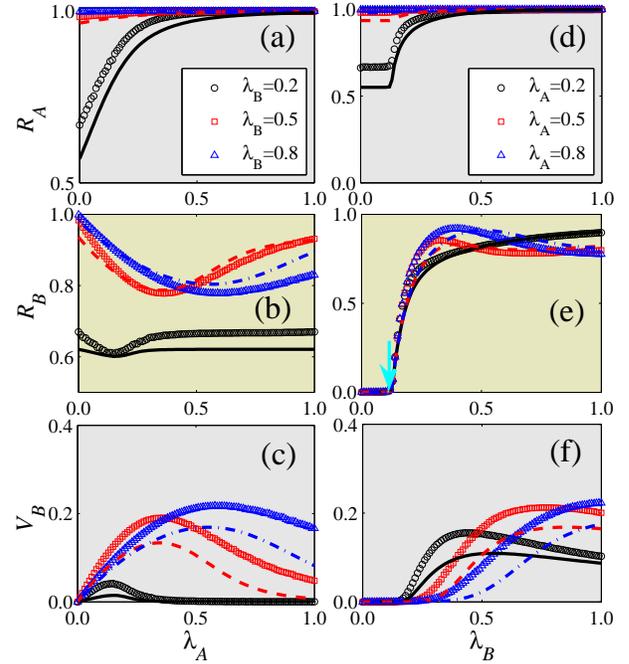,width=1\linewidth}
\caption{(Color online) \textbf{With
      disease transmission rate $\lambda_\mathcal{B}$ being the
      parameter of interest, the asymmetrically interacting dynamics
      spreads on ER-ER networks.} (a) The final information size
  $R_\mathcal{A}$, (b) the final disease size $R_\mathcal{B}$, and (c)
  the vaccination size $V_\mathcal{B}$ versus the information
  transmission rate $\lambda_\mathcal{A}$ for the disease transmission
  rate $\lambda_\mathcal{B}=0.2,0.5$ and 0.8. For
  $\lambda_\mathcal{A}=0.2,0.5$ and 0.8, (d) $R_\mathcal{A}$, (e)
  $R_\mathcal{B}$ and (f) $V_\mathcal{B}$ as a function of
  $\lambda_\mathcal{B}$.  In the figures, symbols are the simulation
  results and the lines are the theoretical predictions.  In (e), the
  arrow indicates the numerical disease threshold. We set other
  parameters to be $\phi=2$ and $p=0.8$.}
\label{fig4}
\end{center}
\end{figure}

Figures~\ref{fig3}(b) and (e) show that $R_\mathcal{B}$ increases with
$\phi$, since individuals are increasingly reluctant to be immunized as
$\phi$ increases, and this causes $V_\mathcal{B}$ to decrease with
$\phi$ [see Figs.~\ref{fig3}(c) and (f)]. Note that $R_\mathcal{B}$ and
$V_\mathcal{B}$ as a function of $\lambda_\mathcal{A}$ have a
non-monotonic shape for $\phi=2$ and $4$, that $R_\mathcal{B}$
($V_\mathcal{B}$) first decreases (increases) with $\lambda_\mathcal{A}$
and then increases (decreases) with $\lambda_\mathcal{A}$. Thus there is
an optimal information transmission rate $\lambda_\mathcal{A}^O$ at
which $R_\mathcal{B}$ ($V_\mathcal{B}$) reaches its minimum (maximum)
value. Qualitatively this is because a node on layer $\mathcal{B}$ will
be immunized only (i) when its counterpart on layer $\mathcal{A}$ is
informed, and (ii) when the number of its infected neighbors
$n_I^\mathcal{B}$ is larger than $\phi$. For a given
$\lambda_\mathcal{B}$, condition (i) is difficult to fulfill when
$\lambda_\mathcal{A}$ is small and the spread of the information is
slow.  Increasing $\lambda_\mathcal{A}$ allows more nodes to fulfill
condition (i) and allows $V_\mathcal{B}$ ($R_\mathcal{B}$) to increase
(decrease) with $\lambda_\mathcal{A}$. When the value of
$\lambda_\mathcal{A}$ is very large the information spreads so rapidly
that condition (ii) can no longer be satisfied. Thus $V_\mathcal{B}$
decreases with $\lambda_\mathcal{A}$, which enhances the spread of
disease.  The optimal phenomenon is not qualitatively affected by the
recovery rates of information and disease. As shown in
Fig.~\ref{fig3}(e), $R_\mathcal{B}$ versus $\lambda_\mathcal{B}$
displays a non-monotonic shape for $\phi=2$ and $4$, i.e.,
$R_\mathcal{B}$ first increases with $\lambda_\mathcal{B}$ and then
decreases. When $\lambda_\mathcal{A}=0.5$ the information spreading is
rapid. Increasing $\lambda_\mathcal{B}$ allows more nodes to fulfill the
second immunization condition and to be immunized [see
  Fig.~\ref{fig3}(f)], and further leads to the decrease ($\phi=2$) or
saturation ($\phi=4$) of $R_\mathcal{B}$ with $\lambda_\mathcal{B}$. The
theoretical predictions of our heterogeneous mean-field theory agree
with the simulation predictions. The differences between the theoretical
predictions and the simulations are caused by the dynamic correlations
among the states of the neighbors and by finite-size network effects
\cite{Wang2014}. The dynamic
    correlations are produced when the information (disease)
    transmission events to one node in layer $\mathcal{A}$
    ($\mathcal{B}$) coming from two distinct neighbors are correlated
    \cite{Altarelli2014}. In the case of coevolution dynamics, the
    dynamic correlations are also induced by the counterparts of
    susceptible nodes \cite{Sanz2014}.

For the disease spreading on layer $\mathcal{B}$, the disease threshold
$\lambda_c^\mathcal{B}$ for $\phi=0$ is clearly larger than the
threshold $\lambda_{c0}^\mathcal{B}=1/\langle k_\mathcal{B}\rangle$,
which is the disease threshold without immunization (i.e., $p=0$) [see
  the right arrow in Fig.~\ref{fig3}(e)]. We can determine the numerical
disease threshold by measuring the susceptibility \cite{Ferreira2012} or
variability \cite{Shu2015} (see details in Method).  Note that the
disease threshold $\lambda_c^\mathcal{B}$ for $\phi\geq1$ is the same as
$\lambda_{c0}^\mathcal{B}$, which is consistent with the theoretical
prediction [see Eq.~(\ref{crit_Epi_Obove}) and the left arrow in
  Fig.~\ref{fig3}(e)].  This occurs because individuals choose
immunization only when the number of their infected neighbors is equal
to or greater than $\phi$.  The asymmetrical coevolution mechanisms
proposed in our model may explain why choosing to be immunized during
disease spreading does not affect the disease threshold
\cite{Fisman2014,Ali2013,Bermejo2006}.

\begin{figure}
\begin{center}
\epsfig{file=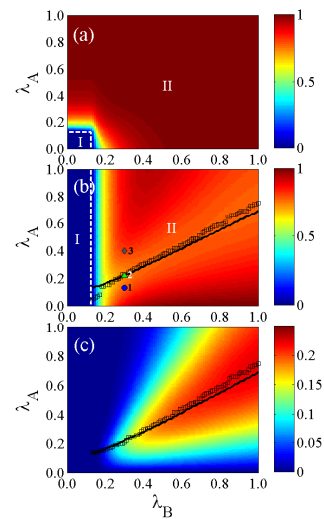,width=0.5\linewidth}
\caption{(Color online) \textbf{Asymmetrically interacting dynamics on
    ER-ER networks.} The final density in each state relating the parameters
  $\lambda_\mathcal{A}$ and $\lambda_\mathcal{B}$: (a) the final
  information size $R_\mathcal{A}$, (b) the final disease size
  $R_\mathcal{B}$ and (c) the vaccination size $V_\mathcal{B}$. In (a),
  the horizontal and vertical dashed lines separate the
  $\lambda_\mathcal{A}-\lambda_\mathcal{B}$ plane into local and global
  information outbreak regions, which are denoted as regions I and
  II. In (b), the vertical dashed line divides the plane into a local
  (region I) and a global (region II) disease outbreak regions. In (b),
  the blue circles ($\lambda_\mathcal{A}=0.13,
  \lambda_\mathcal{B}=0.3$), green up triangle
  ($\lambda_\mathcal{A}=0.22,\lambda_\mathcal{B}=0.3$) and gray diamond
  ($\lambda_\mathcal{A}=0.4,\lambda_\mathcal{B}=0.3$) represent
  $\lambda_\mathcal{A}$ being below, at and above
  $\lambda_\mathcal{A}^O$, respectively (see more discussions in
  Fig.~\ref{fig6}). The black squares (black lines) in (b) and (c)
  represent the optimal information transmission rate
  $\lambda_\mathcal{A}^O$ versus $\lambda_\mathcal{B}$. Other parameters
  are set to be $\phi=2$ and $p=0.8$. }
\label{fig5}
\end{center}
\end{figure}

We use $\phi=2$ to measure the final information and disease sizes (see
Fig.~\ref{fig4}). According to Eq.~(\ref{crit_Epi_Obove}), the disease
threshold is $\lambda_c^\mathcal{B}=1/\langle
k_\mathcal{B}\rangle=0.125$. When $\lambda_\mathcal{B}=0.2$, 0.5, and
0.8, any value of $\lambda_\mathcal{A}$ can cause an information
outbreak due to an outbreak of disease on layer $\mathcal{B}$ [see
  Fig.~\ref{fig4}(a)]. Thus the information outbreak threshold
$\lambda_c^A$ is zero.  Figures~\ref{fig4}(b)--(c) show the optimal
information transmission rate $\lambda_\mathcal{A}^O$ at which
$R_\mathcal{B}$ ($V_\mathcal{B}$) reaches its minimum (maximum)
value. When $\lambda_\mathcal{A} =0.2$, 0.5, and 0.8, $R_\mathcal{A}$
increases with $\lambda_\mathcal{B}$ because of the increase in the
disease [see Fig.~\ref{fig4}(d)].  Note that $\lambda_c^\mathcal{B}$ is
not affected by $\lambda_\mathcal{A}$ [see the arrow in
  Fig.~\ref{fig4}(e)].  As shown in Fig.~\ref{fig4}(e), $R_\mathcal{B}$
versus $\lambda_\mathcal{B}$ first increases and then decreases for
large $\lambda_\mathcal{A} = 0.5$ and 0.8.  This phenomenon can be
understood in the same way with Fig.~\ref{fig3}(e). There is again good
agreement between the theoretical and numerical results.

Figure~\ref{fig5} shows the effects of $\lambda_\mathcal{A}$ and
$\lambda_\mathcal{B}$ on the final steady state for $R_A$, $R_B$, and
$V_B$ for $\phi=2$ and shows the phase diagrams for the final sizes as a
function of $\lambda_A$ and $\lambda_B$. Figure~\ref{fig5}(a) shows that
$R_\mathcal{A}$ increases with $\lambda_\mathcal{A}$ and
$\lambda_\mathcal{B}$. The $\lambda_\mathcal{A}-\lambda_\mathcal{B}$
plane is divided into a local (I) and global (II) information outbreak
regions. In Fig.~\ref{fig5}(a) region I and region II are separated by
the $\lambda_c^\mathcal{A}=1/\langle k_\mathcal{A}\rangle$ (horizontal
white dashed line) and $\lambda_c^\mathcal{A}=1/\langle
k_\mathcal{B}\rangle$ (vertical white dashed line) obtained from
Eq.~(\ref{Lambda_c}). Figure~\ref{fig5}(b) shows how region I and region
II are separated by $\lambda_c^\mathcal{B}$ (see vertical white dashed
line).  For the minimum value of $R_\mathcal{B}$ in region II,
$\lambda_\mathcal{A}^O$ increases linearly with $\lambda_\mathcal{B}$,
as shown in Fig.~\ref{fig5}(b) [see black lines and symbols in (b) and
  (c)]. At the optimal $\lambda_\mathcal{A}^O$, $R_\mathcal{B}$
($V_\mathcal{B}$) reaches its minimum (maximum) value, as shown in
Fig.~\ref{fig5}(b) [Fig.~\ref{fig5}(c)]. Note that
$\lambda_\mathcal{A}^O$ is slightly smaller than $\lambda_\mathcal{B}$
because whether information induces an individual to be vaccinated
depends on the infection level of their neighbors. Our heterogeneous
mean-field theory describes this phenomenon very well.

\begin{figure}
\begin{center}
\epsfig{file=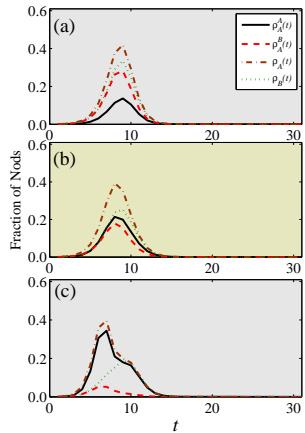,width=0.5\linewidth}
\caption{(Color online) \textbf{On ER-ER coupled networks, the time
    evolution of each type of nodes.} The time evolution of
  $\rho_\mathcal{A}^\mathcal{A}(t)$, $\rho_\mathcal{A}^\mathcal{B}(t)$,
  $\rho_\mathcal{A}(t)$ and $\rho_\mathcal{B}(t)$ for (a)
  $\lambda_\mathcal{A}=0.13$, (b) $\lambda_\mathcal{A}=0.22$ and (c)
  $\lambda_\mathcal{A}=0.40$.  Other parameters are set to be
  $\lambda_\mathcal{B}=0.3$, $\phi=2$ and $p=0.8$.}
\label{fig6}
\end{center}
\end{figure}

Thus we know that for a given disease transmission rate there is an
optimal information transmission rate at which the disease spreading is
markedly reduced. In order to determine the coevolution characteristics
of information and disease spreading when the information reaches its
optimal transmission, we first look at the macroscopic coevolution of
the two dynamics under different information transmission rates as shown
in Fig.~\ref{fig6}.  We denote the fraction of nodes on layer
$\mathcal{A}$ informed by their neighbors or by their counterpart nodes
using $\rho_\mathcal{A}^\mathcal{A}(t)$ and
$\rho_\mathcal{A}^\mathcal{B}(t)$, respectively. Here
$\rho_\mathcal{A}(t)$ [$\rho_\mathcal{B}(t)$] is the fraction of nodes
obtaining the information (disease) on layer $\mathcal{A}$
($\mathcal{B}$) at time $t$.  For small $\lambda_\mathcal{A}=0.13$ below
$\lambda_\mathcal{A}^O$ [see Fig.~\ref{fig6}(a)],
$\rho_\mathcal{A}^\mathcal{A}(t)$, $\rho_\mathcal{A}^\mathcal{B}(t)$,
and $\rho_\mathcal{B}(t)$ reach their peaks simultaneously. Note that
$\rho_\mathcal{B}(t)$ is larger than $\rho_\mathcal{A}^\mathcal{A}(t)$
and very close to $\rho_\mathcal{A}^\mathcal{B}(t)$, which means that
the spread of information is primarily induced by the disease
outbreak. At $\lambda_\mathcal{A}^O=0.22$, we find that
$\rho_\mathcal{A}^\mathcal{A}(t)$, $\rho_\mathcal{A}^\mathcal{B}(t)$,
and $\rho_\mathcal{B}(t)$ reach their peaks simultaneously, and that
$\rho_\mathcal{B}(t)$ is closer to $\rho_\mathcal{A}^\mathcal{A}(t)$
than to $\rho_\mathcal{A}^\mathcal{B}(t)$. Thus the information and
disease have a similar spreading velocity. For a large value of
$\lambda_\mathcal{A}=0.4$, the information spreads more quickly than the
disease. Our results suggest that information and disease spreading have
a similar macroscopic coevolution characteristic when the information
transmission rate is at its optimal value.

\begin{figure}
\begin{center}
\epsfig{file=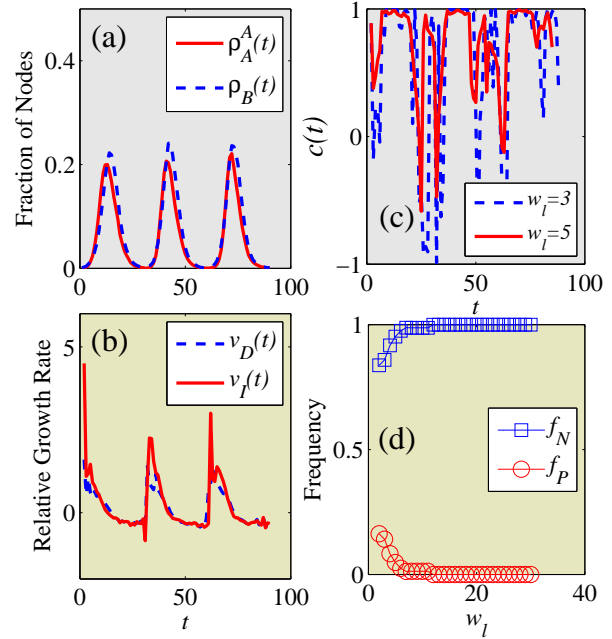,width=1\linewidth}
\caption{(Color online) \textbf{Asymmetrically interacting spreading
    dynamics on coupled ER-ER networks at the optimal information
    transmission rate.} (a) The fractions of nodes in the informed state
  $\rho_\mathcal{A}(t)$ (red solid line) and infected state
  $\rho_\mathcal{B}(t)$ (blue dashed line) versus $t$. (b) The relative
  growth rates $v_D(t)$ (blue dashed line) and $v_I(t)$ (red solid line)
  of $\rho_\mathcal{B}(t)$ and $\rho_\mathcal{A}(t)$ versus $t$,
  respectively.  (c) Cross-correlations $c(t)$ between $v_I(t)$ and
  $v_D(t)$ for the given window size $w_l =3$ (blue dashed line) and
  $w_l=5$ (red solid line). (d) The fractions of negative correlations
  $f_P$ (blue squares) and positive correlations $f_N$ (red circles) as
  a function of $w_l$. We set other parameters to be
  $\lambda_\mathcal{A}=0.22$, $\lambda_\mathcal{B}=0.3$ and $p=0.8$,
  respectively. }
\label{fig7}
\end{center}
\end{figure}

Figure~\ref{fig7} shows the microscopic coevolution characteristics of
the two dynamics at the optimal information transmission
rate. Figure~\ref{fig7}(a) shows the time evolution of information and
disease in three independent dynamical realizations that have similar
trends in their macroscopic coevolution of information spreading and
disease spreading.  Figure~\ref{fig7}(b) shows the relative growth rates
of information $v_I(t)$ and disease $v_D(t)$.  As in the real-world case
in Fig.~\ref{fig1}(b), the same and opposite growth trends are
observed. Figure~\ref{fig7}(c) shows the calculated cross-correlations
between the two time series of $v_D(t)$ and $v_I(t)$. Both positive and
negative cross-correlations exist when the window size is small [see
  Fig.~\ref{fig7}(d)]. Note that Fig.~\ref{fig7} agrees well with the
real-world situation shown in Fig.~\ref{fig1}. Through
    extensive simulations, we find that heterogeneous networks display a
    similar phenomenon. Thus the coevolution between information and
disease can become optimal in which the macroscopic and microscopic
coevolution characteristics of information and disease exhibit similar
trends and the information diffusion greatly suppresses the spread of
disease.

\begin{figure}
\begin{center}
\epsfig{file=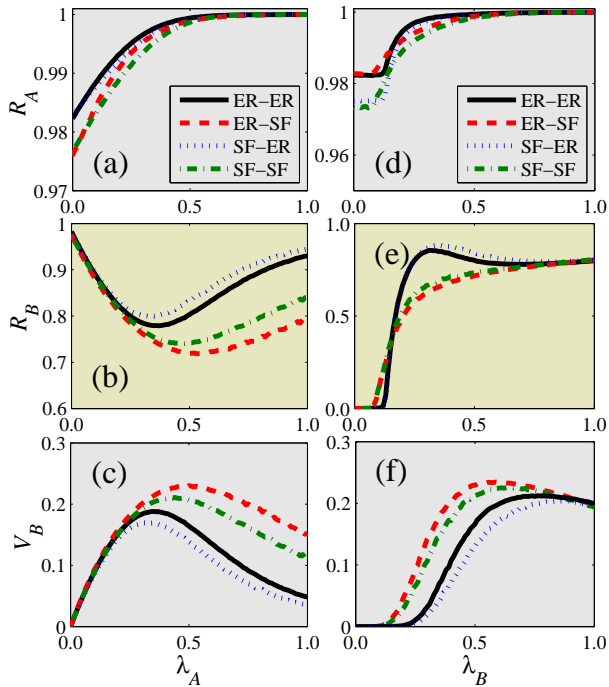,width=1\linewidth}
\caption{(Color online) \textbf{Effect of degree heterogeneity on
    coevolution dynamics}.  (a) The final information size
  $R_\mathcal{A}$, (b) the final disease size $R_\mathcal{B}$ and (c)
  the vaccination size $V_\mathcal{B}$ versus the information transmission
  rate $\lambda_\mathcal{A}$ on ER-ER, ER-SF, SF-ER and SF-SF
  coupled networks with $\lambda_\mathcal{B}=0.5$. For
  ER-ER, ER-SF, SF-ER and SF-SF networks with $\lambda_\mathcal{A}
  =0.5$, (d) $R_\mathcal{A}$, (e) $R_\mathcal{B}$ and (f)
  $V_\mathcal{B}$ as a function of $\lambda_\mathcal{B}$.
  Other parameters are set to be $\phi=2$,
   $p=0.8$ and $\langle k_A\rangle =\langle
  k_B\rangle=8$.}
\label{fig8}
\end{center}
\end{figure}

To examine how topology affects multiplex systems, we next simulate
different possible heterogeneities in the communication and contact
networks (see Fig.~\ref{fig8}). We generate
    scale-free (SF) networks with a power-law degree distribution
    $P(k)\sim k^{-\gamma_D}$ by using an uncorrelated configuration
    model \cite{Catanzaro2005,Yang2012} in which $\gamma_D$ is the
    degree exponent. Through extensive simulations we find that the
    values of $\gamma_D$ do not qualitatively affect the
    results. Without loss of generality we set $\gamma_D=3.0$. Note
that there is an optimal information transmission rate at which the
disease is significantly suppressed [see Figs.~\ref{fig8}(b)--(c)], and
thus heterogeneity in network topology does not qualitatively affect
this optimal phenomenon.  We also find that the multiplex networks with
a homogeneous communication layer and a heterogeneous contact layer have
a greater optimal information transmission rate. As the information
(disease) spreads more (less) widely on homogeneous (heterogeneous)
networks for a large transmission rate, $R_\mathcal{B}$ is further
reduced.  Figure~\ref{fig8}(e) shows that the disease threshold
$\lambda_c^\mathcal{B}$ is determined only by the topology of layer
$\mathcal{B}$, and that the topology of layer $\mathcal{A}$ does not
affect $\lambda_c^\mathcal{B}$.

For information spreading on layer $\mathcal{A}$ as shown in
Fig.~\ref{fig8}(a), $R_\mathcal{A}$ decreases with the degree
heterogeneity of layer $\mathcal{B}$, since a homogeneous contact
network facilitates the spread of disease for large
$\lambda_\mathcal{B}=0.5$ \cite{Pastor-Satorras2001}.  In
Figs.~\ref{fig8}(b)-(c), the effects of the heterogeneity of layer
$\mathcal{A}$ on $R_\mathcal{B}$ and $V_\mathcal{B}$ are negligible when
$\lambda_\mathcal{A}$ is small, but $R_\mathcal{B}$ increases with the
heterogeneity of layer $\mathcal{A}$ when $\lambda_\mathcal{A}$ is large
because it is more difficult to immunize nodes [i.e., $V_\mathcal{B}$
  decreases with the heterogeneity of layer $\mathcal{A}$ in
  Fig.~\ref{fig8}(c)].

Figures~\ref{fig8}(d)--(f) show $R_\mathcal{A}$, $R_\mathcal{B}$ and
$V_\mathcal{B}$ as a function of $\lambda_B$ on several networks for
large $\lambda_\mathcal{A}=0.5$.  The degree heterogeneity of layer
$\mathcal{A}$ is a factor.  When
$\lambda_\mathcal{B}\leq\lambda_c^\mathcal{B}$, $R_\mathcal{A}$
decreases with the heterogeneity of layer $\mathcal{A}$, but the effects
of the heterogeneity of layer $\mathcal{A}$ on $R_\mathcal{B}$ and
$V_\mathcal{B}$ are negligible.  When
$\lambda_\mathcal{B}>\lambda_c^\mathcal{B}$ the heterogeneity of layer
$\mathcal{A}$ does not increase information diffusion, but promotes
disease spreading because nodes are less likely to be immunized. We
examine the effects of the heterogeneity of layer $\mathcal{B}$ and find
that $R_\mathcal{A}$ and $R_\mathcal{B}$ increase (decrease) with the
degree heterogeneity of layer $\mathcal{B}$ for small (large)
$\lambda_\mathcal{B}$. When the degree heterogeneity of layer
$\mathcal{B}$ is increased, the network has a large number of
individuals with very small degrees and more individuals with large
degrees. When $\lambda_\mathcal{B}$ is small there are more hubs in
heterogeneous networks that facilitate disease spreading because they
are more likely to be infected, and this increases information
diffusion. When $\lambda_\mathcal{B}$ is large, however, there are many
small-degree nodes with a low probability of being infected, and this
produces smaller values of $R_\mathcal{B}$, which causes smaller values
of $R_\mathcal{A}$.

\section*{Discussion} \label{sec:dis}

We have systematically investigated the coevolution dynamics of
information and disease spreading on multiplex networks.  We first
discover indications of asymmetrical interactions between the two
spreading dynamics by analyzing real data, i.e., the weekly time series
of information spreading and disease spreading in the form of
influenza-like illness (ILI) evolving simultaneously in the US during an
approximate 200-week period from 3 January 2010 to 10 December
2013. Using these interacting mechanisms observed in real data, we
propose a mathematical model for describing the coevolution spreading
dynamics of information and disease on multiplex networks. We
investigate the coupled dynamics using heterogeneous mean-field theory
and stochastic simulations. We find that information outbreaks can be
triggered by the spreading dynamics within a communications network and
also by disease outbreaks in the disease contact network, but we also
find that the disease threshold is not affected by information
spreading, i.e., that the outbreak of disease is solely dependent on the
topology of the contact network. More important, for a given rate of
disease transmission we find that there is an optimal information
transmission rate that decreases the disease size to a minimum value,
and the modeled evolution of information and disease spreading is
consistent with real-world behavior. We also verify that heterogeneity
in network topology does not invalidate the results.  In addition, we
find that when information diffuses slowly, the degree heterogeneity of
the communication network has a trivial impact on disease spreading. The
homogeneity of the communication network can enhance the vaccination
size and thus prevent disease spreading more effectively when the spread
of information is rapid.

The asymmetrical interacting
    mechanism we discover by analyzing real-world data provides solid
    evidence supporting the basic assumptions of previous researches
    \cite{Granell2013,Wang2014}. Our data-driven model also reveals
some fundamental coevolution mechanisms in the coevolution
dynamics. Using these coevolution dynamics of information and disease we
are able to identify phenomena that differ qualitatively from those
found in previous research on disease-behavior systems. Our results
enable us to quantify the optimal level of information transmission that
suppresses disease spreading. The coevolution mechanisms also enable us
to better understand why the disease threshold is unchanged even when
information spreading in some real-world situations undergoes
coevolution.

Further research on disease-behavior systems promises to discover
additional real-world mechanisms that can be used to refine models of
coevolution spreading dynamics. Developing a more accurate theoretical
method is full of challenges because it is difficult to describe the
strong dynamic correlations among the states of neighboring nodes in a
network. If we take dynamical correlations into account, we may be able
to use such advanced theoretical methods as dynamic message-passing
\cite{Karrer2014,Radicchi2015} or pair approximation
\cite{Eames2002,Gross2006}.

\section*{METHODS}
\textbf{Relative growth rates}.  We define the relative growth rates
$v_G(t)$ of $n_G(t)$ and $v_D(t)$ of $n_D(t)$ to be
\begin{equation} \label{rate_1}
\begin{split}
v_G(t)=\frac{n_G(t+1)-n_G(t)}{n_G(t)}
\end{split}
\end{equation}
and
\begin{equation} \label{rate_2}
\begin{split}
v_D(t)=\frac{n_D(t+1)-n_D(t)}{n_D(t)}.
\end{split}
\end{equation}

If $v_G(t)>0$ [$v_D(t)>0$], $n_G(t)$ [$n_D(t)$] shows an increasing
trend at time $t$. If not, $n_G(t)$ [$n_D(t)$] shows a decreasing trend
at time $t$.

\textbf{Variability measure.}  The variability $\chi$~\cite{Shu2015} is
\begin{equation}
\chi=\frac{\sqrt{\langle R_h^2\rangle-\langle R_h\rangle^2}}{\langle R_h\rangle},
\end{equation}
where $R_h$ is the final information size $R_\mathcal{A}$ or disease
size $R_\mathcal{B}$, and $\langle \cdots\rangle$ is the ensemble
averaging.  The value of $\chi$ exhibits a peak at the critical point
at which the thresholds can be computed.


\section*{FIGURE LEGENDS}
\textbf{Weekly outpatient visits and Google Flu
    Trends (GFT) of influenza-like illness (ILI) from 3 January 2010 to and 21
      September 2013 in the United States.} (a) The relative number of
      outpatient visits $n_D(t)/\langle n_D(t)\rangle$ (blue dashed
      line) and relative search queries aggregated in GFT
      $n_G(t)/\langle n_G(t)\rangle$ (red solid line) versus $t$, where
      $\langle n_D(t)\rangle=\sum_{t=1}^{t_{\rm max}}n_D(t)/ t_{\rm max}
      $ and $\langle n_G(t)\rangle=\sum_{t=1}^{t_{\rm max}}n_G(t)/
      t_{\rm max} $, and $t_{\rm max}$ is the number of weeks.
      (b) The relative growth rate $v_D(t)$ (blue
  dashed line) and $v_G(t)$ (red solid line) of $n_D(t)$ and $n_G(t)$
  versus $t$, respectively. (c) Cross-correlation $c(t)$ between the two
  time series of $v_G(t)$ and $v_D(t)$ for the given window size $w_l=3$
  (blue dashed line) and $w_l=20$ (red solid line).  (d) The fraction of
  negative correlations $f_P$ (blue squares) and positive correlations
  $f_N$ (red circles) as a function of $w_l$. In (a), $n_G(t)$ and
  $n_D(t)$ are divided their average values respectively. In (b), the
  circles and squares denote the relative growth rate at $t=53$ and
  $153$, respectively.

\textbf{FIG.2. Illustration of asymmetrical mechanisms
    of information and disease on multiplex networks.} (a) A multiplex
  network is used to represent communication and contact networks, which
  are denoted as layer $\mathcal{A}$ and layer $\mathcal{B}$,
  respectively. Each layer has 5 nodes. (b) The promotion of information
  spreading by disease.  If node $5$ on layer $\mathcal{B}$ is infected,
  its counterpart on layer $\mathcal{A}$ becomes informed. (c) The
  suppression of disease spreading by information diffusion. Node $3$ in
  layer $\mathcal{B}$ becomes vaccination only when: (1) its counterpart
  on layer $\mathcal{A}$ is in the informed state and (2) the number of
  its infected neighbors on layer $\mathcal{B}$ is equal to the
  threshold $\phi=2$.

\textbf{Fig.3. With
      immunization thresholds $\phi$ being the parameter of interest,
      the final sizes of information, disease and vaccination on two
      layer ER-ER multiplex networks.} (a) The final information size
  $R_\mathcal{A}$, (b) the final disease size $R_\mathcal{B}$, and (c)
  the final vaccination size $V_\mathcal{B}$ versus information
  transmission rate $\lambda_\mathcal{A}$ for different values of
  immunization threshold $\phi$ with $\lambda_\mathcal{B}=0.5$.  For
  different values of $\phi$, (d) $R_\mathcal{A}$, (e) $R_\mathcal{B}$
  and (f) $V_\mathcal{B}$ as a function of $\lambda_\mathcal{B}$ at
  $\lambda_\mathcal{A}=0.5$.  The symbols represent the simulation
  results and the lines are the theoretical predictions obtained by
  numerically solving Eqs.~(\ref{s_A})-(\ref{r_A}) and
  (\ref{s_B})-(\ref{v_B}).  In (e), the two arrows respectively indicate
  the numerical disease thresholds for $\phi\geq1$ and $\phi=0$, which
  are obtained by observing $\chi$. Other dynamical parameters are set
  to be $\lambda_\mathcal{B}=0.5$ and $p=0.8$.

\textbf{FIG.4. With
      disease transmission rate $\lambda_\mathcal{B}$ being the
      parameter of interest, the asymmetrically interacting dynamics
      spreads on ER-ER networks.} (a) The final information size
  $R_\mathcal{A}$, (b) the final disease size $R_\mathcal{B}$, and (c)
  the vaccination size $V_\mathcal{B}$ versus the information
  transmission rate $\lambda_\mathcal{A}$ for the disease transmission
  rate $\lambda_\mathcal{B}=0.2,0.5$ and 0.8. For
  $\lambda_\mathcal{A}=0.2,0.5$ and 0.8, (d) $R_\mathcal{A}$, (e)
  $R_\mathcal{B}$ and (f) $V_\mathcal{B}$ as a function of
  $\lambda_\mathcal{B}$.  In the figures, symbols are the simulation
  results and the lines are the theoretical predictions.  In (e), the
  arrow indicates the numerical disease threshold. We set other
  parameters to be $\phi=2$ and $p=0.8$.

\textbf{FIG.5. Asymmetrically interacting dynamics on
    ER-ER networks.} The final density in each state relating the parameters
  $\lambda_\mathcal{A}$ and $\lambda_\mathcal{B}$: (a) the final
  information size $R_\mathcal{A}$, (b) the final disease size
  $R_\mathcal{B}$ and (c) the vaccination size $V_\mathcal{B}$. In (a),
  the horizontal and vertical dashed lines separate the
  $\lambda_\mathcal{A}-\lambda_\mathcal{B}$ plane into local and global
  information outbreak regions, which are denoted as regions I and
  II. In (b), the vertical dashed line divides the plane into a local
  (region I) and a global (region II) disease outbreak regions. In (b),
  the blue circles ($\lambda_\mathcal{A}=0.13,
  \lambda_\mathcal{B}=0.3$), green up triangle
  ($\lambda_\mathcal{A}=0.22,\lambda_\mathcal{B}=0.3$) and gray diamond
  ($\lambda_\mathcal{A}=0.4,\lambda_\mathcal{B}=0.3$) represent
  $\lambda_\mathcal{A}$ being below, at and above
  $\lambda_\mathcal{A}^O$, respectively (see more discussions in
  Fig.~\ref{fig6}). The black squares (black lines) in (b) and (c)
  represent the optimal information transmission rate
  $\lambda_\mathcal{A}^O$ versus $\lambda_\mathcal{B}$. Other parameters
  are set to be $\phi=2$ and $p=0.8$.

\textbf{FIG.6. On ER-ER coupled networks, the time
    evolution of each type of nodes.} The time evolution of
  $\rho_\mathcal{A}^\mathcal{A}(t)$, $\rho_\mathcal{A}^\mathcal{B}(t)$,
  $\rho_\mathcal{A}(t)$ and $\rho_\mathcal{B}(t)$ for (a)
  $\lambda_\mathcal{A}=0.13$, (b) $\lambda_\mathcal{A}=0.22$ and (c)
  $\lambda_\mathcal{A}=0.40$.  Other parameters are set to be
  $\lambda_\mathcal{B}=0.3$, $\phi=2$ and $p=0.8$.

\textbf{FIG.7. Asymmetrically interacting spreading
    dynamics on coupled ER-ER networks at the optimal information
    transmission rate.} (a) The fractions of nodes in the informed state
  $\rho_\mathcal{A}(t)$ (red solid line) and infected state
  $\rho_\mathcal{B}(t)$ (blue dashed line) versus $t$. (b) The relative
  growth rates $v_D(t)$ (blue dashed line) and $v_I(t)$ (red solid line)
  of $\rho_\mathcal{B}(t)$ and $\rho_\mathcal{A}(t)$ versus $t$,
  respectively.  (c) Cross-correlations $c(t)$ between $v_I(t)$ and
  $v_D(t)$ for the given window size $w_l =3$ (blue dashed line) and
  $w_l=5$ (red solid line). (d) The fractions of negative correlations
  $f_P$ (blue squares) and positive correlations $f_N$ (red circles) as
  a function of $w_l$. We set other parameters to be
  $\lambda_\mathcal{A}=0.22$, $\lambda_\mathcal{B}=0.3$ and $p=0.8$,
  respectively.

\textbf{FIG.8. Effect of degree heterogeneity on
    coevolution dynamics}.  (a) The final information size
  $R_\mathcal{A}$, (b) the final disease size $R_\mathcal{B}$ and (c)
  the vaccination size $V_\mathcal{B}$ versus the information transmission
  rate $\lambda_\mathcal{A}$ on ER-ER, ER-SF, SF-ER and SF-SF
  coupled networks with $\lambda_\mathcal{B}=0.5$. For
  ER-ER, ER-SF, SF-ER and SF-SF networks with $\lambda_\mathcal{A}
  =0.5$, (d) $R_\mathcal{A}$, (e) $R_\mathcal{B}$ and (f)
  $V_\mathcal{B}$ as a function of $\lambda_\mathcal{B}$.
  Other parameters are set to be $\phi=2$,
   $p=0.8$ and $\langle k_A\rangle =\langle
  k_B\rangle=8$.

\acknowledgments

\noindent
This work was partially supported by the National Natural
Science Foundation of China under Grants Nos.~11575041
and 61433014, and China Scholarship Council.
L.A.B. thanks ANCyP, Pict 0429/13 and UNMdP for financial support.

\section*{AUTHOR CONTRIBUTIONS}
W. W. and M. T. devised the research project.
W. W. and Q.-H. L  performed numerical simulations.
W. W., S.-M. C., M. T., L. A. B. and H. E. S. analyzed the results.
W. W., Q.-H. L, S.-M. C., M. T., L. A. B. and H. E. S. wrote the paper.

\section*{Additional information}



{\bf Competing financial interests}:
The authors declare no competing financial interests.

\end{document}